# Upper critical field in the gauge model

A. J. Schofield

*The Cavendish Laboratory, University of Cambridge,*
*Madingley Road Cambridge, CB3 OHE, United Kingdom*

(September 14, 1994)

An analysis of the upper critical field in the gauge model is presented. It is shown that a 'strange metallic' normal state has implications for the pairing transition even if the resulting superconducting state is essentially conventional. The gauge model is considered as one example of an unconventional normal state and the optimally doped and overdoped cases are examined. In the optimally doped regime $B_{c2}$ rises linearly with decreasing temperature to an enhanced value at $T=0$ relative to BCS. In the overdoped regime the gauge model predicts Fermi-liquid behavior but there is a weakly temperature dependent interaction. This can give rise to some upward curvature of the upper critical field and an enhanced zero temperature value.

Recent experiments[1,2] on overdoped single layer (essentially low $T_c$) cuprate superconductors have made the upper critical field accessible throughout its temperature range. The results show a number of anomalies when compared to the weak coupling BCS theory: $B_{c2}(T)$ continues to rise sharply even at the lowest temperatures appearing to intercept the $T=0$ axis with a finite slope and at a field much larger than would be expected from the value of $T_c$. This temperature dependence contrasts with BCS theory where $B_{c2}$ saturates when the magnetic length ($l_B = \sqrt{\hbar/eB}$) is smaller than the electron thermal length ($\xi_f = \hbar\beta v_F/2\pi$). (In this paper a factor of $2\pi$ has been included in $\xi_f$ to simplify formulae.) A second order phase transition is always signalled by the divergence of a susceptibility calculated in the high temperature phase at $T_c$, so the temperature dependence of the upper critical field is essentially a property of the *normal* state. Thus a non-Fermi liquid normal state might be expected to have an unconventional temperature dependence of the upper critical field. This paper explores the consequences on the upper critical field of one such model of the normal state—the gauge model. The optimally doped region of this model is studied although the singular nature of the gauge-field interactions restricts the analysis to qualitative conclusions. The overdoped limit of this model is also treated to address the above experiments. The paper is divided into five sections. The semiclassical theory as applied to two dimensions is briefly reviewed. This defines the terms used throughout the paper and enables a discussion of the role of order parameter symmetry. The gauge model is introduced in its various regimes and the optimal and overdoped regimes are treated separately.

## I. CONVENTIONAL THEORY IN 2D

The theory of the upper critical field has recently been reviewed by Rasolt and Tešanović[3]. The basic theory is considered here to address any possible effect of the order parameter symmetry. In real space the gap equation near $T_c$ may be written as

$$\Delta(\boldsymbol{r}) = \frac{1}{\beta}\sum_{\omega_n}\int d^D\boldsymbol{r}'\int d^D\boldsymbol{l} V(\boldsymbol{r}-\boldsymbol{r}') G_{\omega_n}(\boldsymbol{l},\boldsymbol{r}) G_{-\omega_n}(\boldsymbol{l},\boldsymbol{r}')\Delta(\boldsymbol{l}). \quad (1)$$

An interaction of the form $V_{\mathbf{kk}'} = \sum_n v_n \cos n(\theta_{\mathbf{k}} - \theta_{\mathbf{k}'})$ for $\epsilon_{\mathbf{k}}$ and $\epsilon_{\mathbf{k}'}$ within $\hbar\omega_c$ of the Fermi surfaces is a delta function in real space with a width $r_0 \approx v_F/\omega_c$. The angular dependence of the interaction acts over distances smaller than $r_0$ which is smaller than any other length scale in the problem. One may therefore write

$$\Delta_n(\boldsymbol{r}) = \lambda_n \int_{l>r_0} d^D\boldsymbol{l}\,\Delta_n(\boldsymbol{l}) K(\boldsymbol{l},\boldsymbol{r};\beta), \quad (2)$$

where $K(\boldsymbol{l},\boldsymbol{r}) = \beta^{-1}\sum_{\omega_n} G_{\omega_n}(\boldsymbol{l},\boldsymbol{r}) G_{-\omega_n}(\boldsymbol{l},\boldsymbol{r})$ has been introduced as the pair propagator. A short distance cutoff $r_0$ is also included and the details of the interaction on smaller length scales simply renormalize the coupling strength to $\lambda_n$ in each angular momentum channel. It can therefore be seen explicitly that, neglecting the effects of Pauli pair breaking, the symmetry of the order parameter has no qualitative effect on the temperature dependence of the upper critical field. This is true provided $r_0$ is shorter than $l_B$.

To obtain the well known weak coupling results[4], one makes the semiclassical approximation[5] whereby the effect of the magnetic field on the fermion Green's function is included by a gauge-dependent Aharanov-Bohm phase. Landau level quantization is assumed to be negligible due to temperature or lifetime broadening. In the symmetric gauge this allows the pair propagator in a magnetic field to be written as $K(\boldsymbol{r},\boldsymbol{r}';B) = \exp(i\boldsymbol{r}\times\boldsymbol{r}'/l_B^2)K(|\boldsymbol{r}-\boldsymbol{r}'|;0)$. In this gauge there is a simple solution[6] to Eq. (2) given by $\Delta_n(\boldsymbol{r}) = \Delta_n \exp(-r^2/2l_B^2)$ leading to a pairing instability when

$$\frac{1}{2\pi\lambda_n} = \int_{r_o}^{\infty} K(r,\beta)\exp\left(-\frac{r^2}{2l_B^2}\right) r\, dr \ . \qquad (3)$$

This demonstrates explicitly that the upper critical field is a probe of a normal state correlator.

In real space the Matsubara Green's function for free fermions in 2D is

$$G_{\omega}^{f}(r) = \begin{cases} -\dfrac{im}{2\hbar^2} H_0^{(1)}(r\sqrt{k_F^2 + 2ik_F\omega/v_F}) & \text{if } \omega > 0 \ , \\[4pt] +\dfrac{im}{2\hbar^2} H_0^{(2)}(r\sqrt{k_F^2 + 2ik_F\omega/v_F}) & \text{if } \omega < 0 \ , \end{cases}$$
$$(4)$$

where $H_0^{(1)}$ and $H_0^{(2)}$ are Hankel functions.

In the limit $k_F r \gg 1$ the pair propagator is

$$K_{\Omega}^f(r) = \frac{m^2}{2\pi\hbar^4 \beta k_F r} \frac{e^{-|\Omega|r/v_F}}{\sinh(r/\xi_f)} \ , \qquad (5)$$

where the Matsubara frequency dependence is included as it will be needed later. Putting $\Omega = 0$ one obtains the integral equation for the upper critical field

$$1 = N(0)\lambda_n \int_{r_0}^{\infty} \exp\left(-\frac{r^2}{2l_B^2}\right) \frac{1}{\sinh(r/\xi_f)} \frac{dr}{\xi_f} \ , \qquad (6)$$

where N(0) is the density of states. Notice that the upper critical field becomes temperature independent for $l_B \ll \xi_f$. Note also that the equation may be solved in both the zero temperature and the zero magnetic field limits to yield $B_{c2}(0) = (\pi/\gamma)\phi_0 \left(k_B T_c/\hbar v_F\right)^2$ where $\gamma = \exp(0.577)$ and $\phi_0 = h/e$, the magnetic flux quantum.

In the experiments mentioned above neither of these observations appear to hold true. These results depend on the Fermi-liquid form of the normal state propagators. If the pair condensate is appearing from a non-Fermi liquid normal state one might expect a new form for the upper critical field. This paper treats the upper critical field when there is a pairing transition from the normal state of both the overdoped and optimally doped gauge model of spin-charge separation. An alternative approach to the upper critical field in a spin-charge separated normal state has been undertaken by Dias and Wheatley[7] who assume a form for the normal state Green's function based on a 1D model with spin-charge separation.

## II. GAUGE MODEL PHYSICS

In the gauge model[8] the physical electron is decoupled into a fermion which carries the electron spin quantum number and a boson hole that carries the charge, $c_{\boldsymbol{r}\sigma}^{\dagger} = f_{\boldsymbol{r}\sigma}^{\dagger} b_{\boldsymbol{r}}$. Underlying strong correlations between electrons are reflected in the coupling to a gauge field ('fictitious' in the sense that it exists only in this decoupled approach). The scalar part of the gauge field, $\phi$, ensures that the local density is constant $[\rho_f(\boldsymbol{r}) + \rho_b(\boldsymbol{r}) = 1]$ and the vector potential, $\boldsymbol{a}$, ensures the local currents cancel $[\boldsymbol{j}_f(\boldsymbol{r}) + \boldsymbol{j}_b(\boldsymbol{r}) = 0]$. The partition function for the gauge model in the presence of an external electromagnetic field $A$, may be written as a path integral over fermion and boson coherent states as follows

$$\mathcal{Z}[A] = \int \mathcal{D}[\boldsymbol{a},\phi,f,b] e^{-S_0^f - S_0^b - S_{\text{int}}} \ , \qquad (7)$$

where $S_0^f$ and $S_0^b$ are the actions for non-interacting fermions and bosons respectively and the interaction is given by

$$S_{\text{int}} = \int_0^{\beta} d\tau \int d^2 r \ \boldsymbol{j}_f \cdot \boldsymbol{a} + \boldsymbol{j}_b \cdot (\boldsymbol{a} + e\boldsymbol{A}) + i\phi(\rho_f + \rho_b - 1).$$
$$(8)$$

This model has three possible 'normal' state phases: a strange metal phase ($\langle f_{k\uparrow}^{\dagger} f_{-k\downarrow}^{\dagger} \rangle = 0$, $\langle b_0 \rangle = 0$), a spin-gap phase ($\langle f_{k\uparrow}^{\dagger} f_{-k\downarrow}^{\dagger} \rangle \neq 0$, $\langle b_0 \rangle = 0$) and a Fermi-liquid phase ($\langle f_{k\uparrow}^{\dagger} f_{-k\downarrow}^{\dagger} \rangle = 0$, $\langle b_0 \rangle \neq 0$). The superconducting phase has both $\langle f_{k\uparrow}^{\dagger} f_{-k\downarrow}^{\dagger} \rangle \neq 0$ and $\langle b \rangle \neq 0$. Recent work by Ubbens and Lee[9,10] suggests that the spin-gap phase is thermodynamically stable only for the double-layer compounds like YBCO. This work does not treat the spin-gap phase. Nagaosa and Lee[11] treat the Ginzburg-Landau theory of the spin-charge separated system and the their findings are compared with those obtained in this paper.

The electron pair-propagator in the gauge model is constructed from the underlying fermions and bosons:

$$K^e(\boldsymbol{r},\boldsymbol{r}',\beta;B) = \frac{1}{\beta}\sum_{\Omega} K_{\Omega}^{f}(\boldsymbol{r},\boldsymbol{r}',\beta;B_f) K_{-\Omega}^{b}(\boldsymbol{r}',\boldsymbol{r},\beta;B_b).$$
$$(9)$$

The electron pair is made up of a fermion pair moving from $\boldsymbol{r}$ to $\boldsymbol{r}'$ while a boson pair moves in the opposite direction. The boson Matsubara sum ensures that the pairs arrive together in imaginary time. The bosons and fermions feel different applied magnetic fields due to the screening effect of the fictitious gauge field. When calculated at a Gaussian level these fields are found to be[12]

$$B_f = -\frac{\chi_b}{\chi_b + \chi_f} B \ ; \quad B_b = \frac{\chi_f}{\chi_b + \chi_f} B \ , \qquad (10)$$

where $\chi_b$ and $\chi_f$ are the diamagnetic susceptibilities of the boson and fermion systems respectively.

## III. OPTIMALLY DOPED REGIME

In the optimally doped regime the gauge field is massless. The convolution of the fermion and boson pair-propagators should therefore be performed in the presence of the fluctuating gauge field. This is equivalent to constructing the electron Green's function in the gauge model to derive the pair propagator which has not yet been achieved. Instead a form for the upper critical field in the optimally doped regime is derived here by an approximate consideration of the effect of the gauge-field fluctuations. Recent work[13,14] has suggested that fermions in a fluctuating gauge field have Fermi-liquid response functions at $q = 0$. It will therefore be assumed that the form of the fermion pair-propagator in the presence of the gauge field is unchanged from Eq. (5).

The bosons are strongly effected by the scattering from the gauge field. If this were not so, they would condense into the lowest Landau level and lead to an upper critical field that rises with increasing temperature. As it is, the gauge-field fluctuations will wash out the Landau level structure making a semi-classical approximation possible again. As usual with this problem one can make the quasi-static approximation for the gauge-field propagator which leads to the problem of bosons moving in a static, short-range, random magnetic field. We may obtain the boson pair-propagator from the single particle density matrix by comparing the density matrix with the Green's function in terms of the eigenstates of the problem

$$\rho(r, \beta) = \sum_k \phi_k^*(0) \phi_k(r) e^{-\beta \epsilon_k}, \tag{11}$$

$$G^b_{\omega_n}(r) = \sum_k \frac{\phi_k^*(0) \phi_k(r)}{(i\hbar \omega_n - \epsilon_k + \mu_b)}, \tag{12}$$

$$= -\int_0^\infty \rho(r, \beta) e^{(\mu_b + i\hbar \omega_n)\beta} d\beta. \tag{13}$$

The Laplace transform is well defined for bosons because $\mu_b < 0$. By interchanging the order of summation over Matsubara frequencies and performing the Laplace transforms, one obtains an expression for the boson pair-propagator in terms of the single-particle density matrix

$$K^b_\Omega(r) = \int_0^\infty dx \sum_{m=-\infty}^\infty \rho(r, x) \rho(r, x + |m|\beta)$$
$$\times \exp[(2\mu_b + i\Omega)x + |m|\mu_b \beta]. \tag{14}$$

Since the frequency dependence of the fermion pair-propagator is known from Eq. (5), the frequency convolution of Eq. (9) may be performed yielding the electron pair-propagator

$$K^e(r) = \int_0^\infty dx \sum_{m=-\infty}^\infty \frac{N(0) g(x) \rho(r, x) \rho(r, x+|m|\beta)}{2\pi \xi_f r \sinh(r/\xi_f)} e^{2x\mu_b + |m|\beta}. \tag{15}$$

where

$$g(x) = \frac{1}{\beta} \left[ \frac{e^{r/\xi_f} - \cos(2\pi x/\beta)}{\cosh(r/\xi_f) + \cos(2\pi x/\beta)} - 1 \right]. \tag{16}$$

In the limit that $r \ll \xi_f$ it can be seen that $g(x) \sim \sum_n \delta(x - n\beta)$. This is the relevant limit when the boson length-scale (to be defined later) is smaller than $\xi_f$ and so controls the behavior of the electron pair-propagator.

To make further progress it is necessary to find a form for the single-particle density matrix for bosons in the strange metal phase of the gauge model. While this is not available, an approximate mapping[15] of the single particle problem to the Caldeira-Leggett model of dissipation[16] may be exploited. This mapping can be summarized as follows. The problem of particle motion in the quasi-static limit of gauge-field fluctuations is characterized by the Amperean action

$$S_{\text{Amp}} = S_0 - \frac{1}{4\pi \hbar^2 \chi \beta} \int_0^{\hbar \beta} d\tau \int_0^{\hbar \beta} d\tau' \dot{r}_\tau \ln |r_\tau - r_{\tau'}| \dot{r}_{\tau'}, \tag{17}$$

where $S_0$ is the free particle action and $\chi = \chi_f + \chi_b$ is approximately the Landau diamagnetic susceptibility of the fermions ($1/12\pi m_f$). The best Gaussian action that has the same scaling properties on transformations of the path may be found by substituting the periodic free-particle diffusion law into the log. This leads to an action with the same form as that of the Caldeira-Leggett model

$$S_{\text{CL}} = S_0 - \frac{m}{2\pi \hbar \tau_0} \int_0^{\hbar \beta} d\tau \int_0^{\hbar \beta} d\tau' \dot{r}_\tau \ln \sin \frac{\pi |\tau - \tau'|}{\hbar \beta} \dot{r}_{\tau'},$$
$$= \frac{mR^2}{2\hbar} \left[ \frac{1}{\hbar \beta} + \frac{\ln 2}{\pi \tau_0} \right] + \frac{\beta m}{2} \sum_n \left[ \omega_n^2 + \frac{|\omega_m|}{\tau_0} \right] r_n r_{-n}. \tag{18}$$

Here a general path $r_\tau$ that moves a distance $R$ in imaginary time has been parameterized as $R\tau/\hbar\beta + \sum_n r_n e^{i\omega_n \tau}$ where $\omega_n$ is a boson Matsubara frequency. The form of the density matrix for bosons in the gauge model will be approximated by the single particle density matrix in the above Caldeira-Leggett model. (See Ref. 17 for a critical analysis of this approach.)

The single-particle density matrix in the Caldeira-Leggett model in 2D is found by integrating over the fluctuations from the straight line path to give

$$\rho_{\text{CL}}(r, \beta) = \frac{m_b}{\pi \hbar^2 \beta} \left[ \frac{\Gamma(\hbar \beta \omega_c / 2\pi + 1) \Gamma(\hbar \beta / 2\pi \tau_0 + 1)}{\Gamma(\hbar \beta \omega_c / 2\pi + \hbar \beta / 2\pi \tau_0 + 1)} \right]^2$$
$$\times \exp \left[ -\frac{m_b r^2}{2\hbar} \left( \frac{1}{\hbar \beta} + \frac{\ln 2}{\pi \tau_0} \right) \right]. \tag{19}$$

Here $\omega_c$ is a high-frequency cutoff. A simplified density matrix which captures the asymptotic forms of the free-particle limit ($\hbar\beta \ll \tau_0$) and the dissipative regime ($\hbar\beta \gg \tau_0$) is

$$\rho_{\text{CL}}(r,\beta) \simeq \frac{m_b}{\pi\hbar}\left(\frac{1}{\hbar\beta}+\frac{1}{\tau}\right)\exp\left[-\frac{r^2}{2}\left(\frac{m_b}{\hbar^2\beta}+\frac{1}{\xi_b^2}\right)\right] . \tag{20}$$

Diffusion in the dissipative regime is essentially cutoff by a localization length $\xi_b = \sqrt{\pi\hbar\tau_0/m_b \ln 2}$. A shift in the band edge has been absorbed into the chemical potential. The chemical potential is found from the implicit relation $n_b = \sum_n \rho(0, n\beta) e^{n\mu_b \beta}$.

Working in the dissipative regime $\hbar\beta \gg \tau_0$ with $\xi_b \ll \xi_f$ the electron pair-propagator is found to be

$$K^e(r;\beta) \simeq \frac{N(0)n_b^2}{2\pi\xi_f r}\frac{\exp\left(-r^2/\xi_b^2\right)}{\sinh(r/\xi_f)} . \tag{21}$$

The effective magnetic fields seen by the fermions and bosons are included via the semiclassical approximation. Since these effective fields of Eq. (10) combine to reconstruct the original electron, one may insert the above pair propagator directly into Eq. (3) to obtain the upper critical field which has the form

$$B_{c2}(T) = \frac{\phi_0 m_b \ln 2}{\pi^2 \hbar}\left[\tau_0^{-1}(T_c) - \tau_0^{-1}(T)\right] . \tag{22}$$

Note that the scattering rates that appear are those due to scattering from the gauge-field fluctuations and not that from impurity scattering which may dominate the resistivity at the lowest temperatures.

The mapping to the Caldeira-Leggett model gives the lifetime due to scattering from the gauge field as $\tau_0 = \hbar\beta m_b/3\pi m_f$ (which leads to a physical resistivity linear in $T$). Thus for reasonable values of the boson to fermion mass ratio it is justified to be working in the dissipative regime. Similarly $\xi_b/\xi_f \sim \sqrt{k_B T/\epsilon_F} \ll 1$ so it is clear that the boson length-scale controls the electron pair-propagator. A form for the upper critical field may be now be found

$$B_{c2}(T) = \phi_0 \frac{3\ln 2}{\pi}k_F\left[\frac{k_B(T_c-T)}{\hbar v_F}\right] . \tag{23}$$

Thus we have a linear increase in upper critical field as the temperature is lowered, rising to an enhanced value of $B_{c2}(T = 0)$ compared with that predicted by BCS. The zero-temperature upper critical field is enhanced in this model by a factor $\sim 2k_F\xi_0$ where $\xi_0 = \hbar v_F/\pi\Delta$, the superconducting coherence length. This number is of order 10 in YBCO for example. There is no saturation of the upper critical field within this approach. However, this approximation assumes that the bosons never properly Bose condense but that ring exchange loops continue to grow linearly with $\beta$ rather than exponentially. The effect of Bose condensation is treated in the next section.

## IV. OVERDOPED REGIME

The above discussion has necessarily been qualitative due to the difficulty of studying bosons interacting with a gauge field. These problems are reduced if one works at temperatures and doping regimes where the bosons have condensed. One might expect this to be the case for the overdoped single crystals recently measured. The Bose condensate will screen the gauge field so there is no long-range interaction present. There are two possible mechanisms in this model which could give rise to the unusual temperature dependence of the upper critical field seen experimentally. One is the effect of the convolution of the fermion pair-propagator with the boson pair-propagator in a strictly two-dimensional Bose condensate. The point here is that one normally assumes that once the bosons have condensed their field operators may be replaced by $c$-numbers and conventional Fermi-liquid behavior is recovered. In 2D one can not make this replacement and physical responses may differ from those of the Fermi liquid because of the power law correlations in the Bose condensate and the frequency convolution with this boson response. A second distinct mechanism which may lead to an unconventional upper critical field is the screening of the gauge-field interaction which acquires a temperature dependence due to the temperature dependence of the effective Bose condensate density.

The boson pair-propagator in a 2D superfluid is determined using the method of Popov[18]. A boson repulsion term $u$ is introduced to precipitate a true 2D superfluid state. Such a term arises naturally in the gauge-model description of the $t$-$J$ model. Popov's method involves integrating out the 'fast' fluctuations of the boson field $\psi$ —the fields which fluctuate above some cutoff wavevector $k_0$. This leaves an effective theory expressed in terms of the 'slow' fields. The starting action for the bosons is

$$S = \int_0^{\beta\hbar} d\tau \int d^2\boldsymbol{x}\,\overline{\psi}\partial_\tau\psi - \frac{\hbar^2}{2m}\nabla\overline{\psi}\nabla\psi + \mu\overline{\psi}\psi$$
$$+ \int_0^{\beta\hbar} d\tau \int d^2\boldsymbol{x}\int d^2\boldsymbol{x}' \frac{u(\boldsymbol{x}-\boldsymbol{x}')}{2}\overline{\psi\psi'}\psi'\psi , \tag{24}$$

where the primed fields carry coordinates $(\boldsymbol{x}',\tau)$ and the unprimed fields carry $(\boldsymbol{x},\tau)$. The slow fields are conveniently re-expressed in terms of their density fluctuations $\pi$ [measured from the value at the cutoff $\rho(k_0)$] and their phase variations $\varphi$

$$\psi_{\text{slow}}(\boldsymbol{x},\tau) = \frac{1}{\sqrt{\beta L^2}}\sum_{\omega, k<k_0} e^{i(\omega\tau+k\cdot x)}b(\boldsymbol{k},\omega)$$
$$= [\pi(\boldsymbol{x},\tau) - \rho(k_0)]\exp[-i\varphi(\boldsymbol{x},\tau)] . \tag{25}$$

Here $b$ is the boson annihilation operator. The integration of the fast variables is performed by Popov and he shows that the quadratic part of the effective action of the slow variables may be expressed as

$$S_{\text{eff}} = \sum_{p=(k,\omega)} (\varphi_p \ \pi_p) \begin{pmatrix} \frac{\rho_0 k^2}{m_b} & -\omega \\ \omega & t_0 + \frac{k^2}{4m_b\rho_o} \end{pmatrix} \begin{pmatrix} \varphi_{-p} \\ \pi_{-p} \end{pmatrix} .$$
(26)

Here $\rho_0$ is the boson density and $t_0$ is the $t$-matrix which contains the effect of the fast fields to quadratic order on the slow fields. One now uses this effective action to determine the boson pair-propagator in the 2D superfluid. One ignores the effect of density fluctuations and retains the phase variations to calculate

$$K^{\text{sb}}(r,\tau) = \langle \psi(\boldsymbol{r},\tau)\psi(\boldsymbol{r},\tau)\overline{\psi}(\mathbf{0},0)\overline{\psi}(\mathbf{0},0)\rangle ,$$
$$\approx \rho^2(k_0) \exp -8 \sum_{p=\omega,k<k_0} \langle \varphi_p \varphi_{-p}\rangle \sin^2(\boldsymbol{k}\cdot\boldsymbol{r}+\omega\tau) , \qquad (27)$$

$$\simeq \rho^2 \exp\left[-\frac{m_b v_s}{2\pi\hbar\rho_s r}\left(1+\frac{v_s^2\tau^2}{2r^2}\right)\right]\left[\frac{r}{2\hbar\beta v_s}\right]^{-\frac{m_b}{\pi\hbar^2\beta\rho_s}} . \quad (28)$$

The speed of sound for the Bogoliubov spectrum has been denoted by $v_s = \sqrt{t_0\rho_s/m_b}$ and $\rho_s$ is the superfluid density at $T=0$. This correlator is the square of Popov's result for the Green's function but now generalized to include the imaginary time dependence. However one can see that in the large distance limit the pair propagator becomes static. Finally one obtains the result for the frequency dependence of the superfluid boson pair-propagator

$$K^{\text{sb}}_\Omega(r) = n_b^2 \delta(\Omega) \left(\frac{r}{2\hbar\beta v_s}\right)^{-\frac{m_b}{\pi\hbar^2\beta\rho_s}}$$
$$= n_b^2 \left(r/\xi_s\right)^{-T/T_b} \delta(\Omega) . \qquad (29)$$

The temperature $T_b = \pi\hbar^2\delta/k_B m_b a_0^2$ is the bare boson degeneracy temperature which is much larger than $T_c$ in the overdoped regime. (The number density of bosons has been expressed in terms of the doping fraction $\delta$ and the lattice constant $a_0$.) A new length scale has entered $\xi_s = 2\hbar\beta v_s$, the thermal length for long wavelength phonons in the 2D superfluid. In order to provide an order of magnitude estimate for this, note that $t_0$ is not well defined in 2D as $k \to 0$ so but one may redefine it in terms of the chemical potential[19]. An estimate for the velocity of sound in the dilute limit is $v_s \approx \sqrt{4\pi\hbar^2\delta/m_b a_0^2}$ which, since the fermion and boson masses are of similar order, will be comparable with the Fermi velocity. Thus the pair propagator decays with a very weak power and is delta functionlike in Matsubara frequency. One can conclude that the convolution of the superfluid boson pair-propagator with the fermions does nothing to change the Fermi-liquid form of Eq. (5).

This shows that including a strictly 2D Bose condensate, the form of the upper critical field is unchanged from the Fermi-liquid result in the gauge model. However this omits the residual effect of the screened gauge-field fluctuations. These fluctuations represent a screened transverse current-current interaction between the fermions

$$V_{sg} = \frac{1}{L^2} \sum_{k,k',q} \frac{k^\alpha \text{P}_\text{T}^{\alpha\beta}(q) k'^\beta}{4m_f^2 \rho_s(T)/\hbar^2 m_b} f^\dagger_{k+\frac{q}{2}} f^\dagger_{k'-\frac{q}{2}} f_{k'+\frac{q}{2}} f_{k-\frac{q}{2}} ,$$
(30)

where $\text{P}_\text{T}^{\alpha\beta}(q) = \delta^{\alpha\beta} - q^\alpha q^\beta/q^2$ is the transverse projector. Unlike the 'strange metal' phase where the current-current interaction is singular for low momentum transfers, here the presence of the Bose condensate gives mass to the gauge field and means all momentum transfers are important. To understand the effect of this interaction on the pairing, consider the scattering of two particles on opposite sides of the Fermi surface around an angle $\theta$. One finds an interaction

$$V_{k_F k_F} = \frac{m_b \epsilon_F}{4m_f \rho_s(T)}(1-\cos 2\theta) , \qquad (31)$$

which is repulsive in the s-channel and attractive in the d-channel and in particular leads to a repulsive pseudo-potential $\mu = m_b n_f / 4m_f \rho_s$ for s-wave pairing. By assuming a form for the boson superfluid density one may estimate the effect of this temperature dependent interaction on the upper-critical field. A worst case scenario is that $\rho_s(T) \sim n_b(1-T/T_b)$ which is the intermediate temperature form for $\rho_s(T)$ in the dilute limit found by Fisher and Hohenberg[19]. Figure (1) shows the consequences of such a temperature dependent coupling assuming a Fermi velocity of $1.5 \times 10^5$ m/s, a coupling $N(0)\lambda$ such that $T_c = 15.35$K and a doping of $\delta \sim 0.3$. An effective boson mass of around $6m_f$ and $T_b \sim 500K$ gives an enhancement of $B_{c2}(T=0)$ appropriate to that reported in Ref. 1. This also gives upward curvature of $B_{c2}$ with decreasing temperature but does not explain the dramatic low temperature behavior of $B_{c2}$ seen by Mackenzie *et al.*

The above discussion assumes an instantaneous pairing interaction. If the pairing mechanism is phonon-dominated for these overdoped materials then one should also worry about the effects of a retarded interaction which considerably reduces the effect of the Coulomb pseudo-potential in conventional metals[20,21]. As usual $\mu^* = \mu/[1+\mu\log(\omega_p/\omega_c)]$, where $\omega_p$ is the frequency cut-off scale of the fictitious gauge field. With the bosons condensed this frequency scale will be determined by the condensate density: $\omega_p \approx \hbar\delta/m_b a_0^2$. Thus $\log(\omega_p/\omega_c) \approx \log(T_b/T_D) \approx 2$. While the value of the log is small compared with that for a conventional metal, it is sufficient to remove the unconventional temperature dependencies of the upper critical field seen in Fig. (1).

## V. CONCLUSION

In conclusion, the upper critical field has been studied here as a probe of the normal state pair-propagator. Independent of the symmetry of the order parameter, the Fermi-liquid result shows saturation of the upper critical field at low temperatures. In the gauge model there are at least two possible normal state phases from which a superconducting phase may emerge. In the optimally and low doped phases where there is no spin gap, the upper-critical field is dominated by the short boson thermal length—a consequence of strong scattering from the gauge field. This leads to an enhancement of the zero temperature value for the upper critical field—it is found to be proportional to $T_c$ in contrast to $T_c^2$ for a Fermi liquid. The upper critical field in this regime is found to have a linear temperature dependence with no saturation as long as the bosons remain uncondensed. The slope at $T_c$ will therefore be enhanced relative to BCS in agreement with the phenomenological approach of Nagaosa and Lee[11]. In the overdoped phase it has been shown that the Fermi-liquid temperature dependence should be expected (again consistent with the Ginzburg-Landau approach). However residual scattering from the screened gauge field can lead to some unusual upward curvature of the upper critical field and an enhancement of $B_{c2}(0)$. These effects are almost negligible in the presence of a retarded pairing interaction.

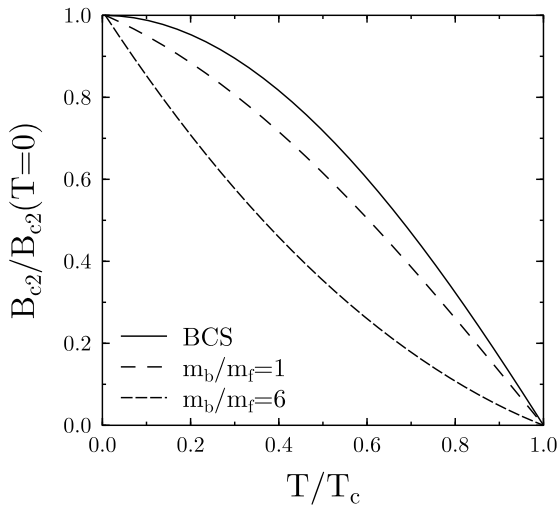

FIG. 1. The normalized upper critical field in the overdoped regime. The BCS result is compared with the effect of a temperature-dependent coupling due to scattering from the screened gauge field. The Bose condensation temperature is $30T_c$ and two values of effective mass ratio are shown. The value of $B_{c2}(0)$ is also enhanced relative to BCS by a factor of 2 for $m_b/m_f = 1$ and 10 for $m_b/m_f = 6$.

## ACKNOWLEDGMENTS

It is a pleasure to acknowledge many important discussions with J. M. Wheatley in particular and also with A. Alexandrov, R. G. Dias, B. Farid and A. P. Mackenzie.